\documentclass[manuscript]{aastex}

\usepackage{emulateapj5}
\usepackage{epsf}
\usepackage{graphics}

\newcommand{\tx}[1]{\textrm{#1}}

\newcommand{\kms}{km\,$\,\tx{s}^{-1}$}

\newenvironment{inlinefigure}{
\def\@captype{figure}
\noindent\begin{minipage}{0.999\linewidth}\begin{center}}
{\end{center}\end{minipage}\smallskip}

\newenvironment{inlinetable}{
\def\@captype{table}
\noindent\begin{minipage}{0.999\linewidth}\begin{center}}
{\end{center}\end{minipage}\smallskip}

\slugcomment{submitted to ApJ Letters}

\shorttitle{The stellar velocity dispersion of MG2016+112}
\shortauthors{Koopmans \& Treu}

\begin{document}

\title{The stellar velocity dispersion of the lens galaxy in
MG2016+112 at z=1.004$^1$} \footnotetext[1]{Based on observations
collected at W.~M. Keck Observatory, which is operated jointly by the
California Institute of Technology and the University of California,
and with the NASA/ESA Hubble Space Telescope, obtained at STScI, which
is operated by AURA, under NASA contract NAS5-26555.}

\author{L\'eon V.E. Koopmans} 
\affil{California Institute of Technology, 
TAPIR, 130-33, Pasadena, CA 91125}
\author{Tommaso Treu}
\affil{California Institute of Technology, 
Astronomy, 105-24, Pasadena, CA 91125}


\begin{abstract}
We present a direct measurement of the stellar velocity dispersion of
the early-type lens galaxy D in the system MG2016+112 ($z=1.004$),
determined from a spectrum obtained with the {\sl Echelle Spectrograph
and Imager} (ESI) on the W.M. Keck--II Telescope, as part of the {\sl
Lenses Structure and Dynamics (LSD) Survey}. We find a velocity
dispersion of $\sigma_{\rm ap}=304\pm27$ km\,s$^{-1}$ inside an
effective circular aperture with a radius of $0.65''$, corresponding
to a central velocity dispersion of $\sigma=328\pm32$
km\,s$^{-1}$. From a {\sl Hubble Space Telescope} F160W--band image,
we measure the effective radius and effective surface brightness in
order to determine the offset of the lens galaxy with respect to the
local Fundamental Plane. The offset corresponds to an evolution of the
rest-frame effective mass-to-light ratio of $\Delta
\log(M/L_B)=-0.62\pm0.08$ from $z=0$ to $z=1.004$. By interpreting
colors and offset of the FP with two independent stellar population
synthesis models, we obtain a single-burst equivalent age of
2.8$\pm$0.8~Gyr (i.e.~$z_{\rm f}$$>$1.9) and a supersolar metallicity
of $\log [Z/Z_\odot]$=0.25$\pm$0.25. The lens galaxy is therefore a
massive elliptical dominated by an old and metal rich stellar
population at $z>1$. The excellent agreement of the stellar velocity
dispersion with that predicted from recent lens models confirms that
the angular separation of the multiple images of the background QSO is
predominantly due to the lens galaxy, and not to a massive ``dark
cluster'', in agreement with recent weak lensing and X--ray
observations. However, the significant overdensity of galaxies in the
field might indicate that this system is a proto-cluster, in formation
around galaxy D, responsible for the $\sim$10\% external shear
inferred from the strong lens models.
\end{abstract}

\keywords{gravitational lensing --- galaxies: elliptical and
lenticular, cD --- galaxies: evolution ---- galaxies: formation ---
galaxies: structure}

\section{Introduction}

In the cold dark matter (CDM) cosmological scenario, structures in the
Universe form through hierarchical merging of smaller structures
(White \& Rees 1978; Blumenthal et al.\ 1984; Davis et al.\ 1985).
Within this general framework, early-type galaxies (E/S0) in the cores
of rich clusters form at high redshifts ($z>2$), corresponding with
the first dark-matter overdensity peaks, as opposed to a later
formation epoch for field E/S0 (Kauffmann 1996).  Clusters
subsequently form around these seeds by accretion of smaller-mass
galaxies, with significant structural and dynamical evolution
occurring at more recent cosmological times. Recent observations have
shown that massive cluster E/S0 were already assembled at $z>1$ and
subsequently evolved passively through ageing of their stellar
populations (e.g. van Dokkum et al.\ 1998; Stanford, Eisenhardt \&
Dickinson 1998). Similarly, field E/S0 galaxies seem not to evolve
dramatically between $z=1$ and $z=0$, both in number (Schade et al.\
1999, Im et al.\ 2002) and structural properties (van Dokkum et al.\
2001; Treu et al.\ 1999, 2001b, hereafter T01b), although secondary
episodes of star formation might be frequent at $z<1$ (Menanteau et
al.\ 2001; Treu et al.\ 2002, hereafter T02).  However, most
observational results on the evolution of E/S0 concern the evolution
of their stellar populations and little is known about the evolution
of their internal structure.

To comprehensively quantify the luminosity, color {\sl and} structural
evolution of the stellar mass and of the dark-matter halo of E/S0 as
function of redshift, we are conducting an observational program with
the Echelle Spectrograph and Imager (ESI) on the W.M.~Keck--II
Telescope: the {\sl Lenses Structure and Dynamics (LSD) Survey}. Aim
of the {\sl LSD survey} is to measure the internal kinematics of a
dozen gravitational-lens galaxies up to $z=1$, allowing a powerful
combination of dynamical and lensing constraints on their mass
distribution. The LSD Survey\footnote{see also
http://www.its.caltech.edu/\~{}koopmans/LSD or
http://www.astro.caltech.edu/\~{}tt/LSD}, its main goals and observing
strategy will be described in detail elsewhere (Treu \& Koopmans 2002,
in preparation).

Here we present the first result of the {\sl LSD Survey}, a
measurement of the stellar velocity dispersion of the lens galaxy in
the system MG2016+112 at $z=1.004$. A summary of relevant prior
observations and a new lens model can be found in Koopmans et al.\
(2002) and references therein. The primary lens galaxy (D) in
MG2016+112 is the highest spectroscopically-confirmed redshift lens
galaxy known to date. The suggestion that this lens was embedded in a
massive ``dark cluster'', based on ASCA X-Ray observations (Hattori et
al. 1997), was recently shown to be incorrect by high-resolution
Chandra observations, showing no evidence for hot X-ray gas (Chartas
et al. 2001). On the other hand, deep optical studies that show an
overdensity of at least six bright E/S0 with the same redshift as the
lens galaxy (Benitez et al. 1999; Soucail et al. 2001; Clowe et
al. 2001).  The absence of a significant weak-lensing signal (Clowe et
al. 2001) and the absence of X-ray emission, however, shows that these
galaxies are not associated with a massive evolved cluster.

In the following, we assume for definiteness that the Hubble constant,
the matter density, and the cosmological constant are
H$_0=65$~km\,s$^{-1}$\,Mpc$^{-1}$, $\Omega_{\rm m}=0.3$, and
$\Omega_{\Lambda}=0.7$, respectively.

\section{Observations}

\subsection{Keck Spectroscopy}

We observed MG2016+112 using ESI on the W.M.~Keck--II Telescope during
four consecutive nights (23--26 July, 2001), with a total integration
time of 8.5\,hrs. The seeing was good ($0\farcs6$-$0\farcs8$) and
three out of four nights were photometric. Between each exposure of
1800\,s, we dithered along the slit to allow for a better removal of
sky residuals in the red end of the spectrum.  The slit (20$''$ in
length) was positioned at a position angle of $13^{\circ}$ in order to
include galaxies D and E, and arc~C$'$ (see~Fig.~\ref{plottwo}).

The slit width of $1\farcs25$ yields an instrumental resolution of
30\,\kms\ which is adequate for measuring the stellar velocity
dispersion and remove narrow sky emission lines. Data reduction was
performed using a newly developed IRAF package\footnote{EASI2D,
developed by D.~Sand and T.~Treu; Sand et al. (2002), in prep.}, that
combines the procedures described in Treu et al.\ (1999, 2001a) and
Smith et al.\ (2001) for the treatment of echelle distortions and sky
subtraction. Spectra of G--K giants observed at twilight with a
$0.3''$ slit width were used as stellar templates, after appropriate
smoothing to match the instrumental resolution of the $1.25''$ slit.
The best fit was obtained for a G4-III stellar template, using the
Gauss--Hermite pixel-fitting software on the spectral region around
the G band (Fig.~\ref{plotone}).  This yielded a stellar velocity
dispersion of $\sigma_{\rm ap}=304\pm27$\,km\,s$^{-1}$ inside an
effective circular aperture with radius 0\farcs65. The scatter in the
fits for the different stellar templates was used to estimate the
uncertainties related to template mismatches (see T01a).  The error on
$\sigma_{\rm ap}$ therefore includes both the random error
contributions and the systematic uncertainties due to template
mismatches, continuum fitting, and fitted spectral range.  The
inferred central velocity dispersion is $\sigma=(1.08\pm 0.05)\cdot
\sigma_{\rm ap}$, i.~e. $\sigma = 328\pm32$~km\,s$^{-1}$. The
correction is based on the average ratio (and spread) between the
central stellar velocity dispersion (i.e. measured in a $R_e/8$
aperture) and the dispersion measured from a larger aperture ($\sim 2
R_e$ in our case), as determined from local E/S0 (see T01a for
details). The predicted velocity dispersion from isothermal
gravitational-lens models, i.e.~320--340~\kms\ (Koopmans et al. 2002),
is in excellent agreement with the spectroscopic determination of the
central stellar velocity dispersion (see also Kochanek 1993, 1994 and
Kochanek et al.\ 2000 for a general discussion).  In addition, we
detect [OII] emission from galaxies D and E at nearly identical
redshifts ($z=1.004$; Fig.2).

\begin{inlinefigure}
\begin{center}
\resizebox{\textwidth}{!}{\includegraphics{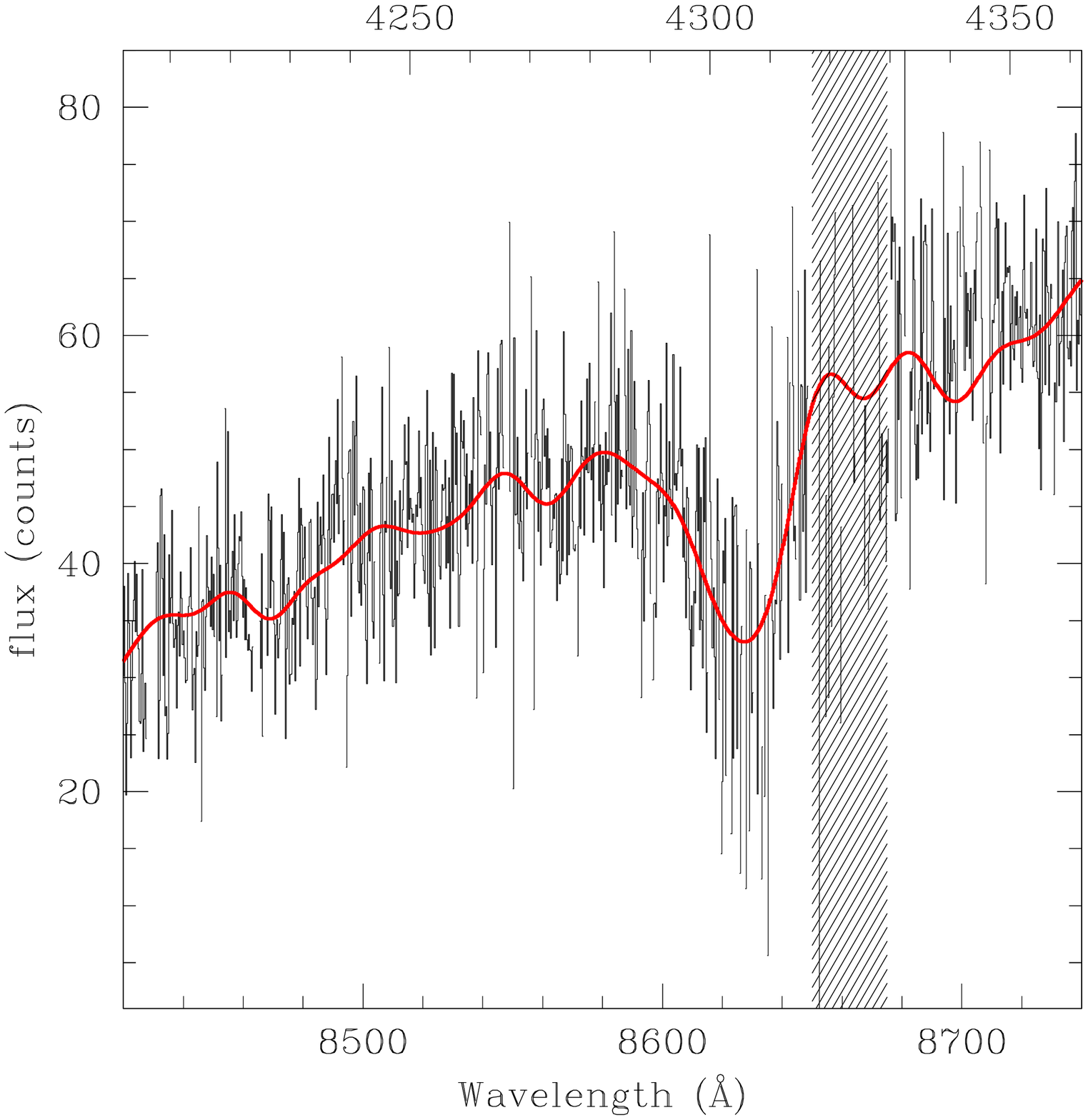}}
\end{center}
\figcaption{The G-band absorption feature at $\sim$$4304$\,\AA\ in the
spectrum of galaxy D in MG2016+112. The rest-frame wavelength scale is
shown on the top axis for reference. The smooth solid line is the
best-fitting template (spectral-type G4--III) convolved to the
measured velocity dispersion $\sigma_{\rm ap}=304\pm27$
km\,s$^{-1}$. The dashed region is affected by sky-subtraction
residuals and was omitted in the fit. \label{plotone}}
\end{inlinefigure}%

\subsection{Hubble Space Telescope Imaging}

Optical and infrared Hubble Space Telescope (HST) images of the system
are available from the HST archive\footnote{Obtained as part of the
CASTLeS Survey}. The Near Infrared Camera and Multi Object
Spectrograph (NICMOS) Camera 2 imaged the galaxy in F160W for 5112s,
while the Wide Field and Planetary Camera 2 imaged the system through
filters F555W and F814W for 5200s each (Fig~\ref{plottwo}).  The
images were reduced and surface photometry performed on the F160W and
F814W images as described in Treu et al.\ (1999, 2001a), except that
cosmic rays rejection on the F814W image was performed using the LA
Cosmic Algorithm (van Dokkum 2001).  The lens galaxy is very faint at
F555W and barely detected in the Planetary Camera images, yielding a
very uncertain flux.  The galaxy brightness profiles are well
represented by an $R^{1/4}$ profile, but the F160W- and F814W--band
effective radii are significantly different and the galaxy shows a
significant reddening towards its core. We attribute the larger
effective radius in F814W--band to a bluer extended envelope around
galaxy~D (seen as a ``wing'' in the brightness profile), which
presumably comes from a younger stellar population that has been
accreted from smaller galaxies (e.g. galaxy E). Because the
F160W--band light is a better tracer of the underlying old stellar
population (dominant in mass) and is not contaminated by the [OII]
emission line (Fig.2), we only use $R_{e,F160W}$ for our analysis. In
addition, the fits provided the integrated magnitudes and effective
surface brightnesses.  The relevant observational quantities of
MG2016+112 are listed in Tab.~1. Rest frame photometric quantities are
corrected for galactic extinction using E(B-V)=0.235 from Schlegel et
al.\ (1998).

\begin{inlinefigure}
\begin{center}
\resizebox{\textwidth}{!}{\includegraphics{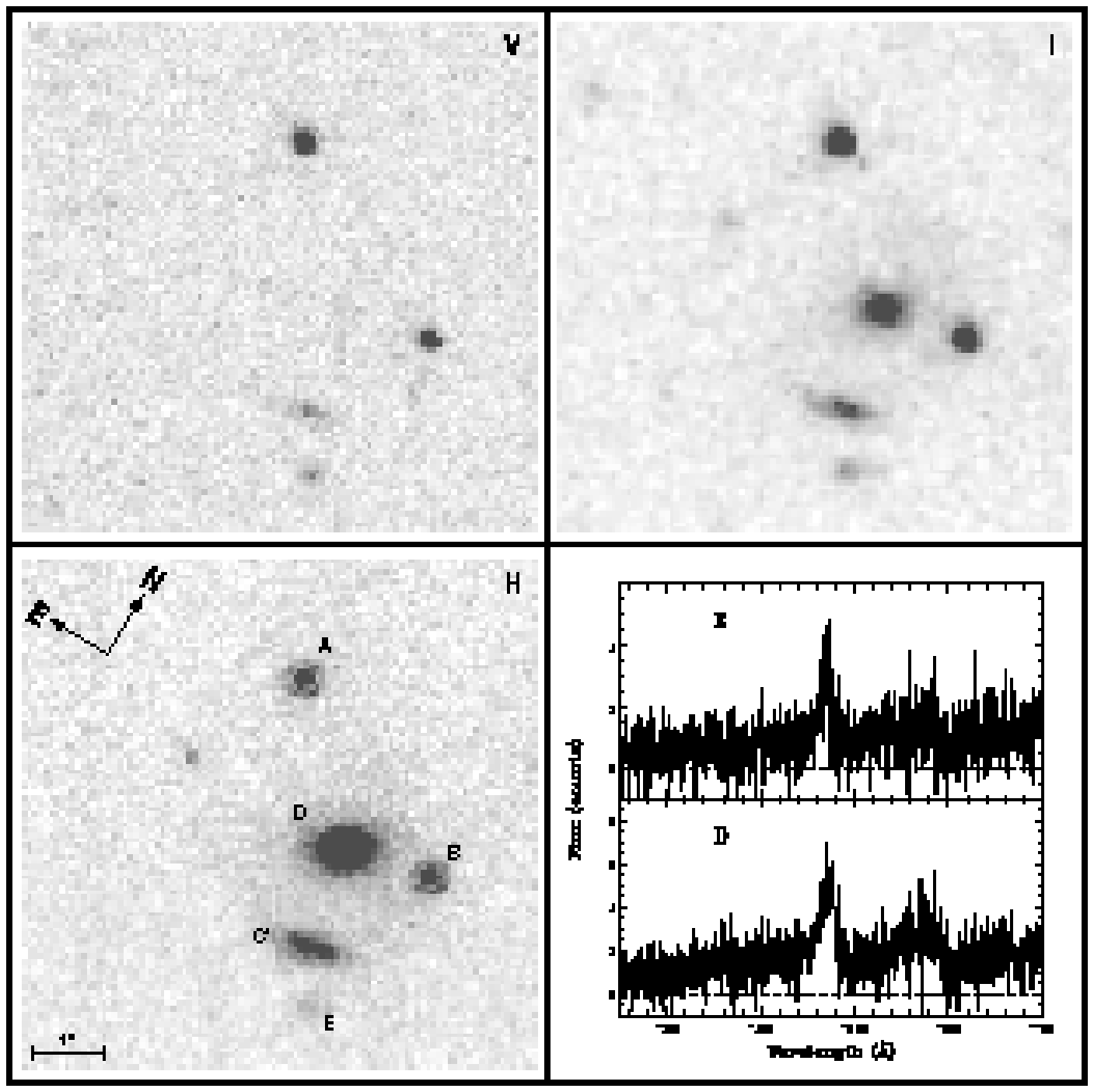}}
\end{center}
\figcaption{HST images of MG2016+112: WFPC2 F555W (V) and F814W (I)
images, the NICMOS-2 F160W (H) image. The exposure times are 5200s (V
and I), and 5120s (H), and the pixels scale are $0.0455''$ (V),
$0.1''$ (I), $0.075''$ (H). The two lensed quasar images (A and B),
the primary lens galaxy (D), the arc C$'$ and the additional
galaxy~(E) are indicated on the H-band image. The emission from the
[OII] doublet detected from galaxies E and D is shown in the bottom
right panel. Note that the doublet in D is blended by the large
velocity dispersion.
\label{plottwo}}
\end{inlinefigure}

\section{Results}

Using the stellar velocity dispersion, effective radius, effective
surface brightness and colors of the lens galaxy, we can now quantify
the evolution of its stellar mass-to-light ratio, as well as the age
and metallicity of the dominant stellar population.  A discussion is
given in Sec.~4.

%
Recent studies have shown that E/S0 in clusters and in the field
define a tight Fundamental Plane (FP) out to $z\sim0.7-0.8$ (van
Dokkum et al.\ 1998; T02), with slopes very similar to the ones
observed in the local Universe. Assuming that galaxy D lies on a FP
with slopes similar to those in the local Universe\footnote{In the
following, we adopt $\log R_e + \log h_{50}= 1.25\,\log \sigma +
0.32\,{\rm SB}_e -8.895$ (Bender et al. 1998) for the local FP in
rest-frame B-band (see T01b for discussion). The uncertainties on the
local determination of the FP are negligible in our analysis.}, we can
obtain the evolution of the intercept $\gamma$ of the FP with
redshift, which is related to the evolution of the average effective
mass to light ratio (M/L). In Fig.~\ref{plotthree} we plot the
evolution of the M/L for cluster and field E/S0 as function of
redshift (small squares and pentagons; from T02), together with the
value obtained for galaxy D (large filled square). From the local FP,
one finds an offset $\Delta \mathrm{SB}_e=-1.55$~mag~arcsec$^2$, which
corresponds to $\Delta \log M/L_B=-0.62\pm0.08$ for pure luminosity
evolution, the error being dominated by the observed FP parameters of
galaxy D. We also note that errors on $R_e$ and SB$_e$ are correlated
and partly cancel in the FP relation (e.g. Treu et al. 2001a; herafter
T01a). The M/L evolution is intermediate between that of clusters and
field E/S0, when we assume that a linear fit, $\Delta\log M/L_B
\propto z$ (see Fig.~\ref{plotthree}), provides a good description of
the data and can be extrapolated to $z=1.004$. Eventhough this is only
a single-galaxy measurement -- not a determination of the FP at $z\sim
1$ -- the inferred M/L evolution suggests that passive evolution of
and old stellar population is a good representation of the evolution
of massive E/S0 out to $z=1.004$. Additional measurements, covering a
range of FP parameters, are needed to confirm this result.

\begin{inlinefigure}
\begin{center}
\resizebox{\textwidth}{!}{\includegraphics{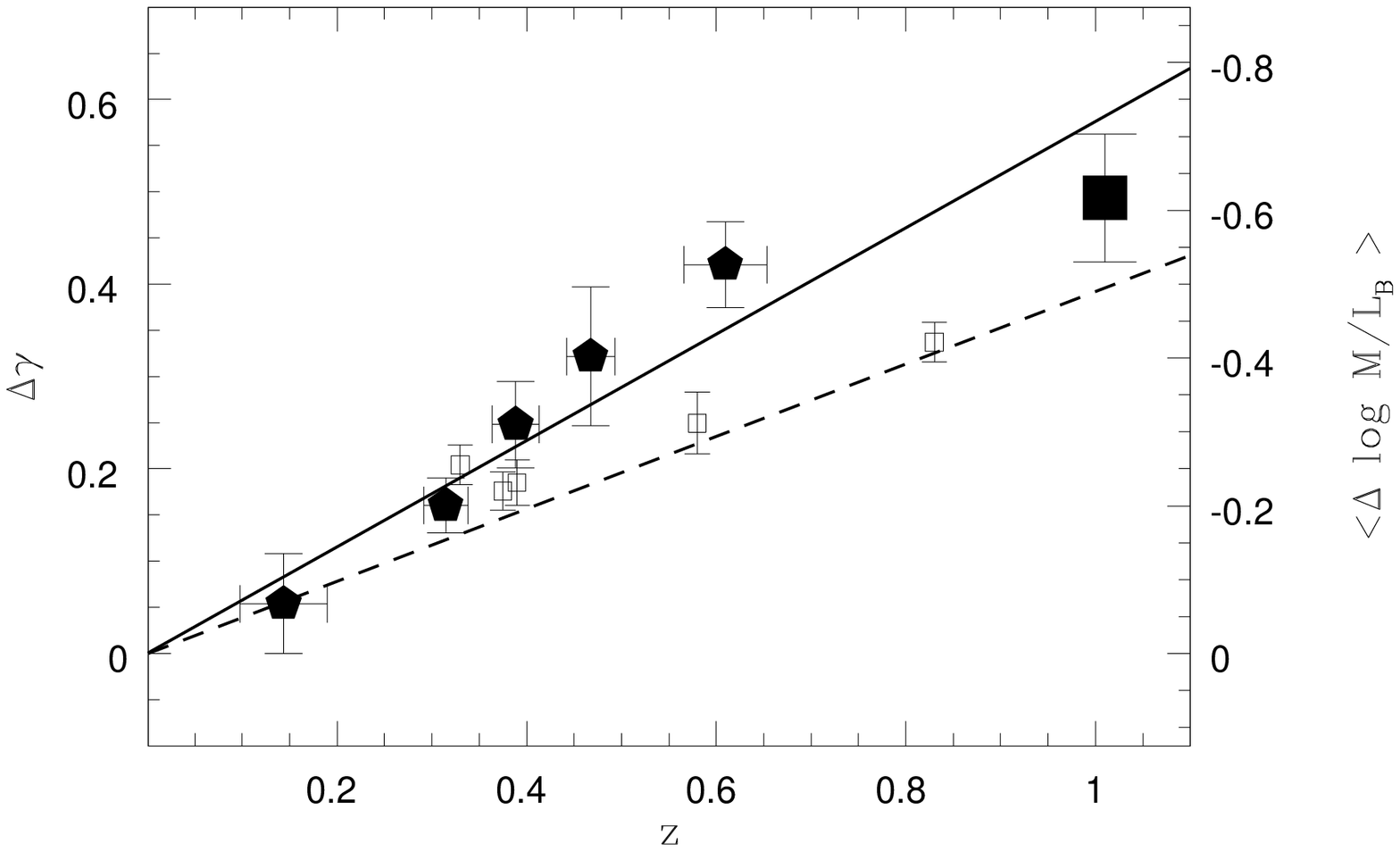}}
\end{center}
\figcaption{Evolution of the M/L as inferred from the evolution of the
FP. Filled pentagons and open squares represent field and cluster
measurements, respectively, while the thick full and dashed lines
indicate linear fits to the M/L evolution for field and cluster E/S0
(see T02 for details). The large filled square at $z=1.004$ indicates
the M/L evolution of galaxy D in MG2016+112.\label{plotthree}}
\end{inlinefigure}

\smallskip

The color and M/L evolution of lens galaxy~D can be used to infer age
and metallicity of its stellar population. A single-burst stellar
population, generated with two independent population synthesis codes
(Bruzual \& Charlot 1993; Fioc \& Rocca-Volmerange 1997), gives a
single-burst equivalent age of 2.8$\pm$0.8~Gyr (i.e.~$z_{\rm
f}$$>$1.9) and a supersolar metallicity
$\log[Z/Z_\odot]$=0.25$\pm$0.25 (Fig.~\ref{plotfour}), assuming there
is no significant internal extinction in the lens galaxy. The contours
in Fig.~\ref{plotfour} are based on the $\chi^2$ difference between
the model and the observed color and mass-to-light ratio evolution
(and their errors).  Because the colors and the luminosity can be
dominated by a small mass-fraction of bright young stars, this age
should be regarded as a lower limit for age of the dominant stellar
population.  The supersolar metallicity is slightly higher, but within
1--$\sigma$ consistent with results obtained from E/S0 in clusters at
redshifts $z\approx 0.3-0.9$, which have been assembled through
hierarchical merging of smaller galaxies (e.g.~Ferreras, Charlot \&
Silk 1999).

\section{Discussion \& Conclusions}

We have presented a direct measurement of the stellar velocity
dispersion of the lens galaxy in the highest
spectroscopically-confirmed redshift lens system MG2016+112
($z=1.004$), as part of the {\sl Lenses Structure and Dynamics (LSD)
Survey}. The effective radius and surface brightness are determined
from HST images, to compare the properties of the lens galaxy with
those of E/S0 in the local Universe. The colors and evolution of
M/L$_B$ of its stellar population, as derived by comparison to the
local FP, indicate an old ($\sim$3\,Gyr) metal-rich stellar
population, consistent with a passively evolved single stellar
population formed at $z_{\rm f}>1.9$.  The large stellar velocity
dispersion and metallicity of the lens galaxy is comparable to the
most massive present-day cluster galaxies (Ferreras et al. 1999;
Fisher et al.\ 1995). In addition, the F814W-band surface brightness
profile at large radii deviates slightly from the $R^{1/4}$ profile,
possibly indicating an outer envelope, as typically found for
brightest cluster galaxies and cD galaxies (Graham et al.\ 1996).
This bluer envelope and the [OII] emission that we detect from galaxy
D (see also Soucail et al. 2001) could be due to the accretion of a
younger stellar populations from smaller gas-rich galaxies, such as
the closest companion to galaxy D for which we detect [OII] in
emission (Fig.2).

\begin{inlinefigure}
\begin{center}
\resizebox{\textwidth}{!}{\includegraphics{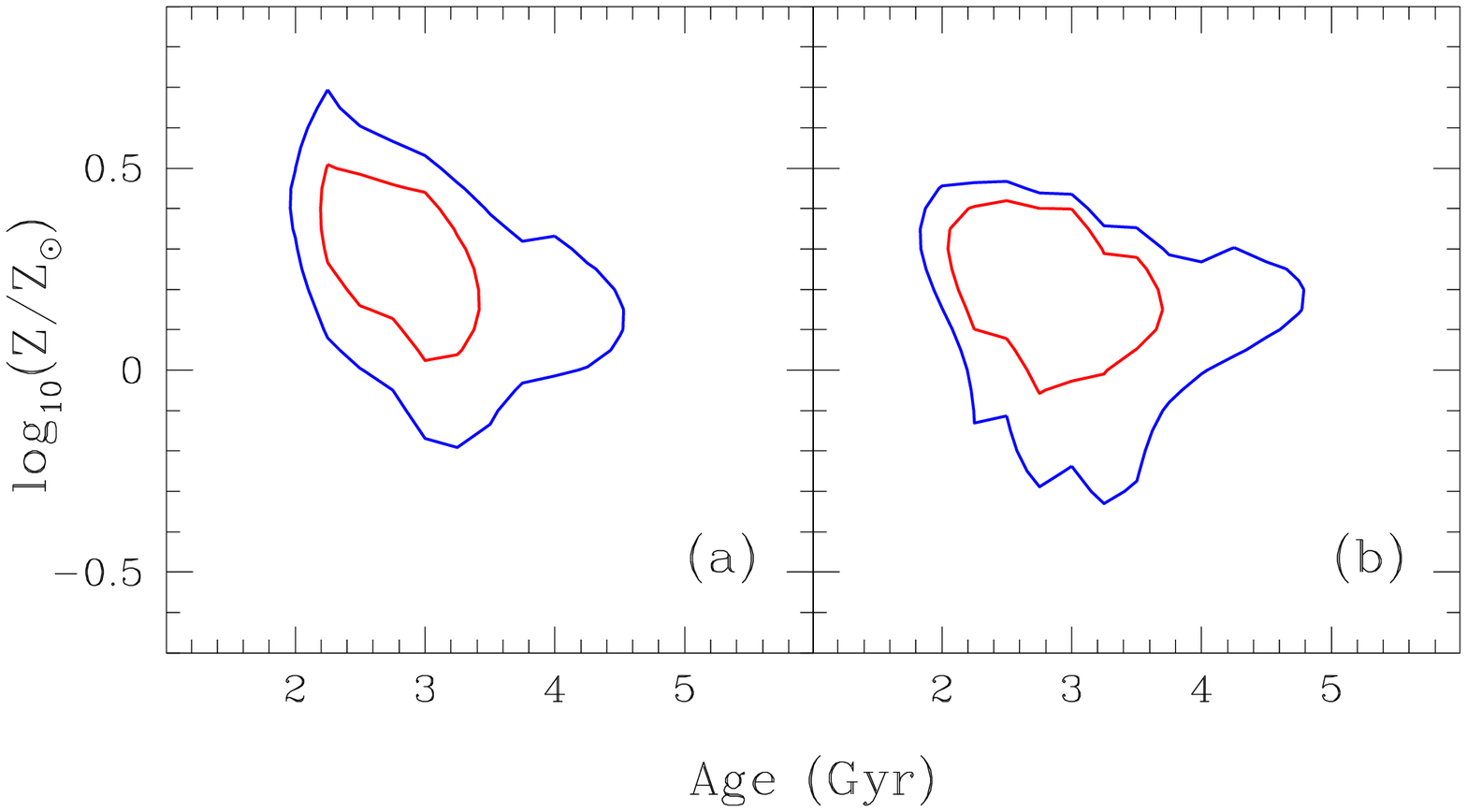}}
\end{center}
\figcaption{Likelihood contours of the age and metalicity of the
stellar population of MG2016+112, galaxy D. The inner and outer
contours indicate the 68\% and 95\% probability. Panels (a) and (b)
are for population synthesis models from Bruzual \& Charlot (1993) and
Fioc \& Rocca-Volmerange (1997), respectively. \label{plotfour}}
\end{inlinefigure}

This scenario is consistent with the gravitational-lens system being
embedded in a galaxy overdensity (i.e. a proto-cluster; see also
Benitez et al.\ 1999), which is not yet relaxed to a more centrally
concentrated system and still shows evidence for ongoing merging and
accretion. In addition, it would explain the absence of a significant
weak-lensing signal (Clowe et al.\ 2001), the relatively small
lensed-image separation, as well as constraints from recent Chandra
X-ray observations (Chartas et al.\ 2001). In particular, the spread
of the velocities of the galaxies in the overdensity (Soucail et al.\
2001), if they are assumed to be virialized, implies a mass three
times larger than the 3--$\sigma$ upper limit inferred from the
non-detection of extended X-ray emission with Chandra (Chartas et
al. 2001).  We therefore think that the assumption of virialization is
{\sl not} correct and that these galaxies form a (non-virialized)
proto-cluster. Three galaxies in the field are spatially clustered at
$z=0.97$ and might be falling towards the overdensity with high
velocity (Soucail et al. 2001), consistent with a cluster in
formation.

~\\
\begin{inlinetable}
\centering
\begin{tabular}{lr}
\hline
\hline
 Redshift (D)          & 1.004$\pm$0.001 \\
 $F160W$ (mag)      & 18.24$\pm$0.02 \\
 $F814W$--$F160W$ (mag)     & 3.3$\pm$0.1 \\
 SB$_{e,F160W}$  (mag/arcsec$^2$)& 17.64$\pm$0.40 \\
 $R_{e,F814W}$ (arcsec) & 0.65$\pm$0.10 \\ 
 $R_{e,F160W}$ (arcsec)& 0.31$\pm$0.06 \\
 $\sigma_*(< 0.65'')$  (km\,s$^{-1}$) &	304$\pm$27 \\
 $b/a$=$(1-e)$ & 0.75$\pm$0.10 \\
 Major axis P.A. ($^\circ$) & $121\pm2$\\
\hline
 $\sigma$  (km\,s$^{-1}$) &	328$\pm$32 \\
 $B-I$ (mag)    & 1.98$\pm$0.15 \\
 $M_{B}$ (mag)  & $-$22.53$\pm$0.10 \\
 $M_{I}$ (mag)  & $-$24.51$\pm$0.04 \\
 SB$_{e,B}$ (mag/arcsec$^2$) & 18.12$\pm$0.50 \\
 SB$_{e,I}$  (mag/arcsec$^2$)& 16.14$\pm$0.40 \\
\hline
\hline
\end{tabular}
\end{inlinetable}

\noindent{Table~1.--- Observed spectro-photometric quantities of
galaxy D in MG2016+112.  The second part of the table lists rest-frame
quantities, derived from the observed quantities (see text).}

{\acknowledgments The use of the Gauss-Hermite Pixel Fitting Software
developed by R.~P.~van der Marel and M.~Franx is gratefully
acknowledged. The ESI data were reduced using software developed in
collaboration with D.~Sand.  We thank E.~Agol, A.~Benson, G.~Bertin,
R.~Blandford, C.~Conselice, R.~Ellis, M.~Stiavelli for comments on
this manuscript. We acknowledge the use of the HST data collected by
the CASTLES collaboration. LVEK and TT acknowledge support by grants
from NSF and NASA (STSCI-AR-09222).}

\clearpage


\begin{thebibliography}{}

\bibitem[Bender et al.(1998)]{1998ApJ...493..529B} Bender, R., Saglia, R.~P., Ziegler, B., Belloni, P., Greggio, L., Hopp, U., \& Bruzual, G.\ 1998, \apj, 493, 529

\bibitem{1999ApJ...527...31B} Benitez N., Broadhurst T., Rosati P.,
Courbin F., Squires G., Lidman C., Magain P., 1999, \apj, 527, 41

\bibitem{B84} Blumenthal, G.~R., Faber S.~M., Primack J.~R. \& Rees, M.~J., 1984, Nature, 311, 517

\bibitem{1993ApJ...405..538B} Bruzual A. G., Charlot S., 1993, \apj, 405,
553 

\bibitem{2001ApJ...550L.163C} Chartas G., Bautz M., Garmire G., Jones
C., Schneider D. P., 2001, \apjl, 550, L167

\bibitem{2001A&A...369...16C} Clowe, D., N.~Trentham, and J.~Tonry
2001, \aap 369, 16-25


\bibitem{1985ApJ...292..371D} Davis, M., Efstathiou, G., Frenk, C. S.,
White, S. D. M., 1985, \apj, 292, 394 

\bibitem{FCS99} Ferreras, I., Charlot, S., Silk, J., 1999, \apj, 521, 81

\bibitem{1997A&A...326..950F} Fioc M., Rocca-Volmerange B., 1997,
\aap, 326, 962

\bibitem{1995ApJ...438..539F} Fisher D., Illingworth G., Franx M.,
1995, \apj, 438, 562

\bibitem{1996ApJ...465..534G} Graham A., Lauer T. R., Colless M., 
Postman M., 1996, \apj, 465

\bibitem{1997Natur.388..146H} Hattori M. et al., 1997, \nat, 388, 148

\bibitem[Im et al.\ 2002]{Im02} Im, M., Faber, S.~M., Koo, D.~C.,
Phillips, A.~C., Schiavon, R.~P., Simard, L. \& Willmer, C.~N.~A.,
2002, ApJ, in press

\bibitem{1996MNRAS.281..487K} Kauffmann, G., 1996, \mnras, 281, 492

\bibitem{1994ApJ...436...56K} Kochanek, C. S., 1994, \apj, 436, 66

\bibitem{2000ApJ...543..131K} Kochanek C. S. et al., 2000, \apj, 543, 148

\bibitem{astro-ph/0106575} Koopmans L.V.E., Garrett M.A., Blandford
R.D., Lawrence C.R., Patnaik A.R., Porcas R.W., 2002, \mnras, accepted

\bibitem[Menanteau, Abraham \& Ellis (2001)]{M01} Menanteau, F.,
Abraham, R.~G., Ellis, R.~S. 2001, MNRAS, 322, 1

\bibitem[Schade et al.\ 1999]{CFRS-Es} Schade D. et al., 1999, ApJ,
525, 31

\bibitem{1998ApJ...500..525S} Schlegel D. J., Finkbeiner D. P., Davis
M., 1998, \apj, 500

\bibitem{S01} Smith, G.~P., Treu, T., Ellis, R.~S., Smail, I.~R.,
Kneib, J.-P., Frye, B.~L., 2001, \apj, 562, 635

\bibitem{2001A&A...367..741S} Soucail G., Kneib J.-P., Jaunsen A. O.,
Hjorth J., Hattori M., Yamada T., 2001, \aap, 367, 747


\bibitem{1998ApJ...492..461S} Stanford, S. A., Eisenhardt, P. R.,
Dickinson M., 1998, \apj, 492 

\bibitem{T02} Treu, T., Stiavelli, M., Casertano, S., M{\o}ller, P.,
Bertin, G., 2002, \apjl, 564, L13

\bibitem{T99} Treu, T., Stiavelli, M., Casertano, S., M{\o}ller, P.,
Bertin, G., 1999, \mnras, 308, 1037

\bibitem{2001MNRAS.326..221T} Treu, T., Stiavelli, M., M{\o}ller, P.,
Casertano, S., \& Bertin, G., 2001a, \mnras, 326, 221

\bibitem{2001MNRAS.326..237T} Treu, T., Stiavelli, M., Bertin, G.,
Casertano, S., M{\o}ller, P., 2001b, \mnras, 326, 237

\bibitem{LACOSMIC} van Dokkum, P.~G., 2001, PASP, 113, 1420

\bibitem{1998ApJ...504L..17V} van Dokkum P. G., Franx M., Kelson
D. D., Illingworth, G. D., 1998, \apjl, 504

\bibitem{1999ApJ...520L..95V} van Dokkum P. G., Franx M., Fabricant
D., Kelson D. D., Illingworth G. D., 1999, \apjl, 520, L98

\bibitem{2001ApJ...552L.101V} van Dokkum, P. G., Stanford, S. A.,
Holden, B. P., Eisenhardt, P. R., Dickinson, M., \& Elston, R., 
2001, \apjl, 552, L104 

\bibitem{WR78} White, S.~D.~M. \& Rees, M.~J., 1978, MNRAS, 183, 341

\end{thebibliography}
\end{document}